\documentclass[aps,pra,twocolumn,groupedaddress,showpacs,floatfix]{revtex4}

\usepackage{graphicx}

\input BoxedEPS.tex
\SetTexturesEPSFSpecial
\HideDisplacementBoxes

\begin{document}

 \title{Dense Atom Clouds in a Holographic Atom Trap}

\author{R. Newell}\author{J. Sebby}\author{T. G. Walker}
\affiliation{Department of Physics, University of Wisconsin-Madison,
		Madison, Wisconsin 53706 }
\date{\today}

\pacs{1234567}

\begin{abstract}
We demonstrate the production of high density cold atom samples ($2\times10^{14}$ atoms/cm$^3$) in a simple optical lattice formed
with YAG light that is diffracted from a holographic phase plate. A loading protocol is described that results in 10,000 atoms per
lattice site.  Rapid free evaporation leads to phase space densities of 1/150 within 50 msec.  The resulting small, high
density atomic clouds are very attractive for a number of experiments, including ultracold Rydberg atom physics.
\end{abstract}

\maketitle

Conventional light-force atom traps, in particular the magneto-optical trap (MOT), are limited in density by radiation trapping to
typically 10$^{11}$ cm$^{-3}$
\cite{Walker90,Ketterle93}.  In the experiments described here, we attain densities in far-off-resonance traps (FORTs)  that
exceed this density by 3 orders of magnitude, and are comparable to BEC densities.  We show
that  FORTs can achieve these densities in tens of milliseconds, as compared to tens of seconds for magnetic traps.  Such rapidly
produced high density sources are of particular interest for studying novel collision phenomena that do not require the coherence
of BECs and yet can be exploited for quantum manipulation and entanglement of atoms. They are also attractive for  evaporative
cooling; several groups have recently cooled atoms to quantum degeneracy using FORTs \cite{Barrett01,Granade02}.

FORTs \cite{Miller93}  use spatially varying intensities $I({\bf r})$ to produce a
conservative potential
$U({\bf r})=-2\pi\alpha(\lambda) I({\bf r})/c$, where $\alpha(\lambda)$ is the polarizability of the atom at the FORT laser
wavelength
$\lambda$. At a given temperature $T$ and number of trapped atoms $N$ the peak density depends not directly on the trap depth but
rather on its curvature:
$n\sim N(\kappa/T)^{3/2}$, where
$\kappa=-\partial^2_{\bf r} U$.  Therefore much higher densities can be obtained at roughly the same trap depth by
forming an interference pattern (optical lattice) to obtain a rapid spatial variation in the intensity.  In general, for a
fixed laser power  and beam area, forming an interference pattern  (lattice) of cell size $\Lambda$ from
$M$ beams  increases the trap depth over a single-beam FORT by a factor $O(M)$ and the spring constant by a
factor $O(M/\Lambda^2)$.  However, the number of atoms per lattice site is reduced by
$O(\Lambda^3)$ so that the geometrical factors cancel and the density scales as $M^{3/2}$.
Thus there is considerable freedom in choosing the geometry of lattice FORTs without sacrificing density.

\begin{figure}[htb]
\includegraphics[scale=0.33]{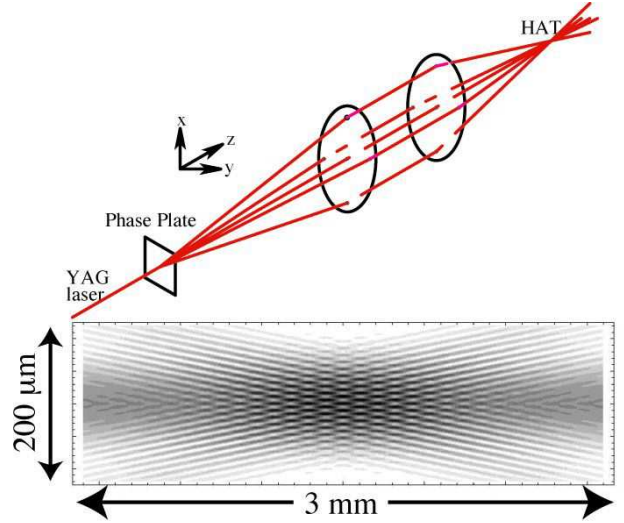}
\caption{Optical train used to produce the HAT, and calculation of the interference pattern formed at the intersection of the five
HAT beams.}\label{HATgeom}
\end{figure}

For phenomena occurring at high densities ($>10^{13}$ cm$^{-3}$) it is desirable to construct a lattice FORT with sufficient unit
cell volume that many atoms can be loaded into a given lattice site. This has been demonstrated by several groups using a variety
of trap geometries:  3D lattice \cite{Han00}, crossed dipole \cite{Barrett01}, a single retro-reflected beam \cite{Friebel98,
Granade02}, a beam focused with an array of microlenses \cite{Dumke02}, and the interference of light diffracted from a holographic
phase plate
\cite{Boiron98}.  All save the last two require interferometric stability and single-mode lasers.   In an early experiment Boiron
{\it et al.} used a holographic phase plate and a YAG laser to construct a large $\Lambda$ Cs lattice. After applying blue Sisyphus
cooling,
 they obtained
densities of order 10$^{13}$ cm$^{-3}$.  In this work we use higher intensities, optimized loading, and evaporation  without the
aid of Sisyphus cooling to attain Rb densities exceeding 10$^{14}$ cm$^{-3}$  and phase space densities approaching 1/150 after
loading for only tens of milliseconds.  Furthermore our use of a multimode flashlamp-pumped laser demonstrates the robust nature of
the trap.

In our experiments we use a YAG laser to acheive an optical lattice, called a Holographic Atom Trap (HAT), by interfering five
diffracted beams from a holographic phase plate \cite{MEMS}  (Fig.~\ref{HATgeom}).  At the intersection of the 5 laser
beams an interference pattern
\begin{equation}
{I({\bf r})\over I_0}=\left|e^{ikz}+2\beta e^{ikz(1-\theta^2/2)}\left(\cos kx\theta +\cos ky\theta\right)\right|^2
\end{equation}
 is produced and   Rb atoms ($\alpha(1.06 \mu {\rm m})=105
 $\AA$^3$\cite{KadarKallen92,Marinescu94b}) are trapped at the intensity maxima.  Here $k=2\pi/\lambda$, $\theta=0.1$ rad is the
diffraction angle of the four first-order beams, $\beta^2=0.20$ is the ratio of the intensity of a single first-order
diffracted beam to that of the zeroth order beam $I_0$. Along the $\hat{z}$ propagation direction of the light the interference
arises from the Talbot effect\cite{Boiron98}.  The resulting lattice sites (microtraps) are 10
$\mu$m$\times$ 10
$\mu$m $\times$ 100 $\mu$m in size. With the individual beams focussed to 80
$\mu$m and a total power of 8 W the  depth of the central microtrap is 500 $\mu$K and the  oscillation frequencies 
%for atoms trapped
%in the microtraps 
are 17 kHz, 17 kHz, and 0.7 kHz.  These calculated frequencies were experimentally confirmed by  parametric
heating of the HAT atoms \cite{Friebel98}.
\begin{figure}[htb]
\includegraphics[scale=0.36]{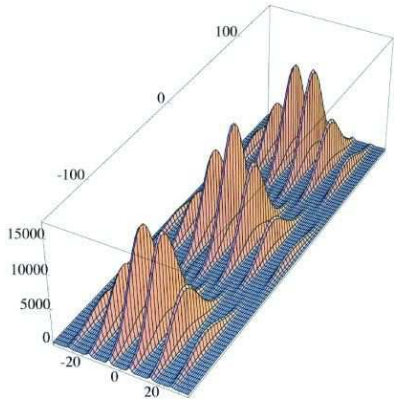}
\caption{Calaculate density distribution of atoms in the HAT, assuming a Boltzmann distribution with a temperature
equal to 1/10 the maximum trap depth.  Distances are in microns, densities in arbitrary units.}\label{HATpot}
\end{figure}
Since the lasers are nearly copropagating, the potential is quite stable against vibrations and the YAG laser can have a large
bandwidth.  Our multi-longitudinal mode laser has a bandwidth of approximately 25 GHz.  Heating due to intensity noise
\cite{Gehm98} was easily eliminated with an acousto-optic intensity stabilizer.  The relatively large lattice sites  allow many
atoms ($\sim 10^4$) to be trapped in each site.

\begin{figure}[htb]
\includegraphics[scale=0.22]{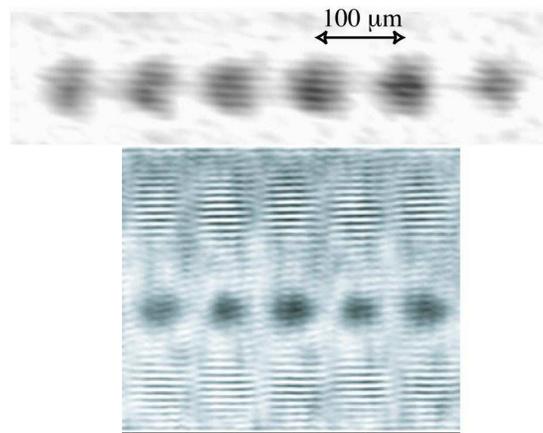}
\caption{Top: Image showing
microtraps.  Bottom: first order Bragg diffraction is shown in the fringes.  }
\label{Bragg}
\end{figure}

\begin{figure}[htb]
\includegraphics[scale=0.6]{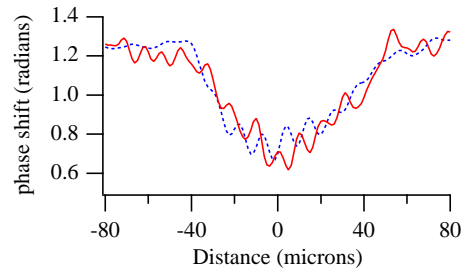}
\caption{Spatial profiles of two neighboring Talbot images.  The microtrap structure is shifted between the two, as expected.}
\label{TalbotProfile}
\end{figure}

Fig.~\ref{HATpot} shows a calculated density distribution for atoms trapped in the HAT at 50 $\mu$K temperature. 
We use a probe laser
propagating along the
$\hat{y}$-direction to take  spatial heterodyne  \cite{Kadlecek01} phase images that show the atoms' isolation within the Talbot
fringes and the microtraps (Fig.~\ref{Bragg}). 
 We  observe Bragg
diffraction of the probe beam from the atoms in the microtraps by interference with the probe beam, also shown in
Fig.~\ref{Bragg}.   An interesting observation from Fig.~\ref{HATpot} is that there should be a relative misalignment of the
microtraps in successive Talbot fringes.  This is confirmed by the spatial profiles shown in Fig.~\ref{TalbotProfile}.
Analysis of such images gives about $4\times10^5$ atoms per Talbot fringe. Time-of-flight temperature measurements of 50
$\mu$K coupled with knowledge of the trapping potential implies peak densities exceeding
$2\times 10^{14}$ cm$^{-3}$ and phase space densities of 1/150.  Typically 25 microtraps are occupied within
each Talbot fringe, with 10\% of the atoms in the central microtrap.

Key to the HAT's success is an efficient loading protocol.  We begin with a vapor cell forced dark spot MOT \cite{Anderson94} with
50\% of the 10$^7$ atoms in each of the hyperfine ground states (5S$_{1/2}$ F=1,2) of $^{87}$Rb.  The  dark spot is acheived by
imaging an opaque object in the hyperfine repumping beam (F=1$\rightarrow$5P$_{3/2}$ F$'$=2).  The trapping light
(F=2$\rightarrow$F$'$=3) is tuned 3 linewidths $\Gamma$ below resonance and has  an intensity of 72 mW/cm$^2$.  We add a
depumping laser, tuned to the high frequency side of the F=2$\rightarrow$F$'$=2 resonance, to optically pump more atoms down to
the F=1 ground state.  To load the HAT from the  MOT, we  compress the cloud by decreasing the trapping light intensity by
a factor of 3, then increasing the MOT magnetic field.  After 20 msec we
turn on the HAT laser.  AC Stark shifts tune the repumping and trapping beams away from resonance and the depumping beam towards
resonance, causing the atoms  in the HAT to be extremely dark (estimated $\sim 0.001$ in F=2 inside the FORT)
\cite{Kuppens00}. During the HAT loading phase we shift the trapping laser detuning to $-9\Gamma$. The number of trapped atoms
increases until it reaches steady state in about 50 msec, after which the MOT lasers are extinguished.  The MOT to HAT transfer
efficiency is as high as 15\%. 

\begin{figure}[tb]
\includegraphics[scale=0.56]{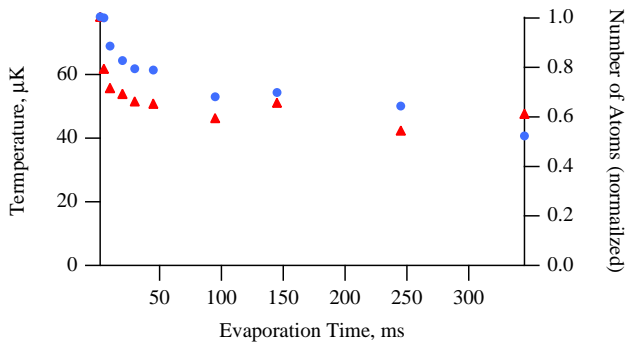}
\caption{Evaporative cooling of the atoms occurs after the cooling lasers are switched off.  The number of atoms in a
single Talbot fringe(circles) and the temperature (triangles) rapidly decrease until the temperature reaches roughly 1/10 the
trap depth.}
\label{evap}
\end{figure}

We show in Fig.~\ref{evap} the temperature and number of atoms vs time  after the loading ceases.  The temperature rapidly
decreases to a value of about 50 $\mu$K, approximately 1/10 of the trap depth,  as expected for free evaporation of atoms from the
HAT \cite{OHara01}.  The evaporation is rapid due to the very high densities; we estimate the elastic collision rate  to exceed
3000/s.  Such high collision rates make evaporative cooling very promising for the HAT and such experiments are underway in our
laboratory. Trap lifetime studies show that after the evaporation ceases the temperature stays fixed but the atoms are slowly
ejected from the trap and/or heated \cite{Bali99} by background gas collisions with a time constant of about 500 msec.  For
comparison, the MOT lifetime is typically 2-3 seconds for our vapor-loaded trap.

These high density samples are of potential interest for a variety of experiments. In addition to obvious examples such as
evaporative cooling and cold collisions, many groups are interested
in ultracold Rydberg atoms. Excitation of Rydberg states of principal quantum number
$\sim$100  in our HAT will give rise to strong, long-range dipole-dipole interactions. Because atoms in our microtraps are
localized within a 15 $\mu$m region, these interaction energies are in excess of 1 MHz for atoms on opposite sides of a
microtrap.    Consequently, excitation by a narrow-band laser of more than one Rydberg atom at a time per microtrap should be
greatly suppressed.  This ``dipole blockade'' is of great interest for applications in quantum information
processing\cite{Lukin01} and single atom and photon sources\cite{Saffman02}.  Similarly, the densities are high enough to expect
efficient excitation of novel long-range Rydberg molecules such as those recently proposed \cite{Greene00,Boisseau02}.  In
addition, it will be interesting to see how the very much higher densities acheived in the HAT as compared to MOTs affect the
production of cold plasmas\cite{Killian01}.

 The authors are grateful for the support of NASA, the NSF, and the ARO. We thank Steve Kadlecek and
Jason Day for pivotal contributions to the imaging system, and David Steele and Nathan Harrison for work on early stages of the
experiment. 

%\bibliography{lasercooling}

\begin{thebibliography}{23}
\expandafter\ifx\csname natexlab\endcsname\relax\def\natexlab#1{#1}\fi
\expandafter\ifx\csname bibnamefont\endcsname\relax
  \def\bibnamefont#1{#1}\fi
\expandafter\ifx\csname bibfnamefont\endcsname\relax
  \def\bibfnamefont#1{#1}\fi
\expandafter\ifx\csname citenamefont\endcsname\relax
  \def\citenamefont#1{#1}\fi
\expandafter\ifx\csname url\endcsname\relax
  \def\url#1{\texttt{#1}}\fi
\expandafter\ifx\csname urlprefix\endcsname\relax\def\urlprefix{URL }\fi
\providecommand{\bibinfo}[2]{#2}
\providecommand{\eprint}[2][]{\url{#2}}

\bibitem[{\citenamefont{Walker et~al.}(1990)\citenamefont{Walker, Sesko, and
  Wieman}}]{Walker90}
\bibinfo{author}{\bibfnamefont{T.}~\bibnamefont{Walker}},
  \bibinfo{author}{\bibfnamefont{D.}~\bibnamefont{Sesko}}, \bibnamefont{and}
  \bibinfo{author}{\bibfnamefont{C.}~\bibnamefont{Wieman}},
  \bibinfo{journal}{Phys. Rev. Lett.} \textbf{\bibinfo{volume}{64}},
  \bibinfo{pages}{408} (\bibinfo{year}{1990}).

\bibitem[{\citenamefont{Ketterle et~al.}(1993)\citenamefont{Ketterle, Davis,
  Joffe, Martin, and Pritchard}}]{Ketterle93}
\bibinfo{author}{\bibfnamefont{W.}~\bibnamefont{Ketterle}},
  \bibinfo{author}{\bibfnamefont{K.}~\bibnamefont{Davis}},
  \bibinfo{author}{\bibfnamefont{M.~A.} \bibnamefont{Joffe}},
  \bibinfo{author}{\bibfnamefont{A.}~\bibnamefont{Martin}}, \bibnamefont{and}
  \bibinfo{author}{\bibfnamefont{D.~E.} \bibnamefont{Pritchard}},
  \bibinfo{journal}{Phys. Rev. Lett.} \textbf{\bibinfo{volume}{70}},
  \bibinfo{pages}{2253} (\bibinfo{year}{1993}).

\bibitem[{\citenamefont{Barrett et~al.}(2001)\citenamefont{Barrett, Sauer, and
  Chapman}}]{Barrett01}
\bibinfo{author}{\bibfnamefont{M.~D.} \bibnamefont{Barrett}},
  \bibinfo{author}{\bibfnamefont{J.~A.} \bibnamefont{Sauer}}, \bibnamefont{and}
  \bibinfo{author}{\bibfnamefont{M.~S.} \bibnamefont{Chapman}},
  \bibinfo{journal}{Phys. Rev. Lett.} \textbf{\bibinfo{volume}{87}},
  \bibinfo{pages}{010404} (\bibinfo{year}{2001}).

\bibitem[{\citenamefont{Granade et~al.}(2002)\citenamefont{Granade, Gehm,
  O'Hara, and Thomas}}]{Granade02}
\bibinfo{author}{\bibfnamefont{S.~R.} \bibnamefont{Granade}},
  \bibinfo{author}{\bibfnamefont{M.~E.} \bibnamefont{Gehm}},
  \bibinfo{author}{\bibfnamefont{K.~M.} \bibnamefont{O'Hara}},
  \bibnamefont{and} \bibinfo{author}{\bibfnamefont{J.~E.}
  \bibnamefont{Thomas}}, \bibinfo{journal}{Phys. Rev. Lett.}
  \textbf{\bibinfo{volume}{88}}, \bibinfo{pages}{120405}
  (\bibinfo{year}{2002}).

\bibitem[{\citenamefont{Miller et~al.}(1993)\citenamefont{Miller, Cline, and
  Heinzen}}]{Miller93}
\bibinfo{author}{\bibfnamefont{J.~D.} \bibnamefont{Miller}},
  \bibinfo{author}{\bibfnamefont{R.~A.} \bibnamefont{Cline}}, \bibnamefont{and}
  \bibinfo{author}{\bibfnamefont{D.~J.} \bibnamefont{Heinzen}},
  \bibinfo{journal}{Phys. Rev. A} \textbf{\bibinfo{volume}{47}},
  \bibinfo{pages}{R4567} (\bibinfo{year}{1993}).

\bibitem[{\citenamefont{Han et~al.}(2000)\citenamefont{Han, Wolf, Oliver,
  McCormick, DePue, and Weiss}}]{Han00}
\bibinfo{author}{\bibfnamefont{D.~J.} \bibnamefont{Han}},
  \bibinfo{author}{\bibfnamefont{S.}~\bibnamefont{Wolf}},
  \bibinfo{author}{\bibfnamefont{S.}~\bibnamefont{Oliver}},
  \bibinfo{author}{\bibfnamefont{C.}~\bibnamefont{McCormick}},
  \bibinfo{author}{\bibfnamefont{M.~T.} \bibnamefont{DePue}}, \bibnamefont{and}
  \bibinfo{author}{\bibfnamefont{D.~S.} \bibnamefont{Weiss}},
  \bibinfo{journal}{Phys. Rev. Lett.} \textbf{\bibinfo{volume}{85}},
  \bibinfo{pages}{724} (\bibinfo{year}{2000}).

\bibitem[{\citenamefont{Friebel et~al.}(1998)\citenamefont{Friebel,
  Scheunemann, Walz, Hansch, and Weitz}}]{Friebel98}
\bibinfo{author}{\bibfnamefont{S.}~\bibnamefont{Friebel}},
  \bibinfo{author}{\bibfnamefont{R.}~\bibnamefont{Scheunemann}},
  \bibinfo{author}{\bibfnamefont{J.}~\bibnamefont{Walz}},
  \bibinfo{author}{\bibfnamefont{T.~W.} \bibnamefont{Hansch}},
  \bibnamefont{and} \bibinfo{author}{\bibfnamefont{M.}~\bibnamefont{Weitz}},
  \bibinfo{journal}{Appl. Phys. B} \textbf{\bibinfo{volume}{B67}},
  \bibinfo{pages}{699} (\bibinfo{year}{1998}).

\bibitem[{\citenamefont{Dumke et~al.}(2002)\citenamefont{Dumke, Volk, Muther,
  Buchkremer, Birkl, and Ertmer}}]{Dumke02}
\bibinfo{author}{\bibfnamefont{R.}~\bibnamefont{Dumke}},
  \bibinfo{author}{\bibfnamefont{M.}~\bibnamefont{Volk}},
  \bibinfo{author}{\bibfnamefont{T.}~\bibnamefont{Muther}},
  \bibinfo{author}{\bibfnamefont{F.~B.~J.} \bibnamefont{Buchkremer}},
  \bibinfo{author}{\bibfnamefont{G.}~\bibnamefont{Birkl}}, \bibnamefont{and}
  \bibinfo{author}{\bibfnamefont{W.}~\bibnamefont{Ertmer}},
  \bibinfo{journal}{Phys. Rev. Lett.} \textbf{\bibinfo{volume}{89}},
  \bibinfo{pages}{097903} (\bibinfo{year}{2002}).

\bibitem[{\citenamefont{Boiron et~al.}(1998)\citenamefont{Boiron, Michaud,
  Fournier, Simard, Sprenger, Grynberg, and Salomon}}]{Boiron98}
\bibinfo{author}{\bibfnamefont{D.}~\bibnamefont{Boiron}},
  \bibinfo{author}{\bibfnamefont{A.}~\bibnamefont{Michaud}},
  \bibinfo{author}{\bibfnamefont{J.~M.} \bibnamefont{Fournier}},
  \bibinfo{author}{\bibfnamefont{L.}~\bibnamefont{Simard}},
  \bibinfo{author}{\bibfnamefont{M.}~\bibnamefont{Sprenger}},
  \bibinfo{author}{\bibfnamefont{G.}~\bibnamefont{Grynberg}}, \bibnamefont{and}
  \bibinfo{author}{\bibfnamefont{C.}~\bibnamefont{Salomon}},
  \bibinfo{journal}{Phys. Rev. A} \textbf{\bibinfo{volume}{57}},
  \bibinfo{pages}{R4106} (\bibinfo{year}{1998}).

\bibitem[{MEM()}]{MEMS}
\bibinfo{note}{MEMS Optical, Huntsville, AL}.

\bibitem[{\citenamefont{Kadar-Kallen and Bonin}(1992)}]{KadarKallen92}
\bibinfo{author}{\bibfnamefont{M.~A.} \bibnamefont{Kadar-Kallen}}
  \bibnamefont{and} \bibinfo{author}{\bibfnamefont{K.~D.} \bibnamefont{Bonin}},
  \bibinfo{journal}{Phys. Rev. Lett.} \textbf{\bibinfo{volume}{68}},
  \bibinfo{pages}{2015} (\bibinfo{year}{1992}).

\bibitem[{\citenamefont{Marinescu et~al.}(1994)\citenamefont{Marinescu,
  Sadeghpour, and Dalgarno}}]{Marinescu94b}
\bibinfo{author}{\bibfnamefont{M.}~\bibnamefont{Marinescu}},
  \bibinfo{author}{\bibfnamefont{H.~R.} \bibnamefont{Sadeghpour}},
  \bibnamefont{and} \bibinfo{author}{\bibfnamefont{A.}~\bibnamefont{Dalgarno}},
  \bibinfo{journal}{Phys. Rev. A} \textbf{\bibinfo{volume}{49}},
  \bibinfo{pages}{5103} (\bibinfo{year}{1994}).

\bibitem[{\citenamefont{Gehm et~al.}(1998)\citenamefont{Gehm, O'Hara, Savard,
  and Thomas}}]{Gehm98}
\bibinfo{author}{\bibfnamefont{M.~E.} \bibnamefont{Gehm}},
  \bibinfo{author}{\bibfnamefont{K.~M.} \bibnamefont{O'Hara}},
  \bibinfo{author}{\bibfnamefont{T.~A.} \bibnamefont{Savard}},
  \bibnamefont{and} \bibinfo{author}{\bibfnamefont{J.~E.}
  \bibnamefont{Thomas}}, \bibinfo{journal}{Phys. Rev. A}
  \textbf{\bibinfo{volume}{58}}, \bibinfo{pages}{3914} (\bibinfo{year}{1998}).

\bibitem[{\citenamefont{Kadlecek et~al.}(2001)\citenamefont{Kadlecek, Sebby,
  Newell, and Walker}}]{Kadlecek01}
\bibinfo{author}{\bibfnamefont{S.}~\bibnamefont{Kadlecek}},
  \bibinfo{author}{\bibfnamefont{J.}~\bibnamefont{Sebby}},
  \bibinfo{author}{\bibfnamefont{R.}~\bibnamefont{Newell}}, \bibnamefont{and}
  \bibinfo{author}{\bibfnamefont{T.~G.} \bibnamefont{Walker}},
  \bibinfo{journal}{Opt. Lett.} \textbf{\bibinfo{volume}{26}},
  \bibinfo{pages}{137} (\bibinfo{year}{2001}).

\bibitem[{\citenamefont{Anderson et~al.}(1994)\citenamefont{Anderson, Petrich,
  Ensher, and Cornell}}]{Anderson94}
\bibinfo{author}{\bibfnamefont{M.~H.} \bibnamefont{Anderson}},
  \bibinfo{author}{\bibfnamefont{W.}~\bibnamefont{Petrich}},
  \bibinfo{author}{\bibfnamefont{J.~R.} \bibnamefont{Ensher}},
  \bibnamefont{and} \bibinfo{author}{\bibfnamefont{E.~A.}
  \bibnamefont{Cornell}}, \bibinfo{journal}{Phys. Rev. A}
  \textbf{\bibinfo{volume}{50}}, \bibinfo{pages}{R3597} (\bibinfo{year}{1994}).

\bibitem[{\citenamefont{Kuppens et~al.}(2000)\citenamefont{Kuppens, Corwin,
  Miller, Chupp, and Wieman}}]{Kuppens00}
\bibinfo{author}{\bibfnamefont{S.~J.~M.} \bibnamefont{Kuppens}},
  \bibinfo{author}{\bibfnamefont{K.~L.} \bibnamefont{Corwin}},
  \bibinfo{author}{\bibfnamefont{K.~W.} \bibnamefont{Miller}},
  \bibinfo{author}{\bibfnamefont{T.~E.} \bibnamefont{Chupp}}, \bibnamefont{and}
  \bibinfo{author}{\bibfnamefont{C.~E.} \bibnamefont{Wieman}},
  \bibinfo{journal}{Phys. Rev. A} \textbf{\bibinfo{volume}{62}},
  \bibinfo{pages}{013406} (\bibinfo{year}{2000}).

\bibitem[{\citenamefont{O'Hara et~al.}(2001)\citenamefont{O'Hara, Gehm,
  Granade, and Thomas}}]{OHara01}
\bibinfo{author}{\bibfnamefont{K.~M.} \bibnamefont{O'Hara}},
  \bibinfo{author}{\bibfnamefont{M.~E.} \bibnamefont{Gehm}},
  \bibinfo{author}{\bibfnamefont{S.~R.} \bibnamefont{Granade}},
  \bibnamefont{and} \bibinfo{author}{\bibfnamefont{J.~E.}
  \bibnamefont{Thomas}}, \bibinfo{journal}{Phys. Rev. A}
  \textbf{\bibinfo{volume}{64}}, \bibinfo{pages}{051403}
  (\bibinfo{year}{2001}).

\bibitem[{\citenamefont{Bali et~al.}(1999)\citenamefont{Bali, O'Hara, Gehm,
  Granade, and Thomas}}]{Bali99}
\bibinfo{author}{\bibfnamefont{S.}~\bibnamefont{Bali}},
  \bibinfo{author}{\bibfnamefont{K.~M.} \bibnamefont{O'Hara}},
  \bibinfo{author}{\bibfnamefont{M.~E.} \bibnamefont{Gehm}},
  \bibinfo{author}{\bibfnamefont{S.~R.} \bibnamefont{Granade}},
  \bibnamefont{and} \bibinfo{author}{\bibfnamefont{J.~F.}
  \bibnamefont{Thomas}}, \bibinfo{journal}{Phys. Rev. A}
  \textbf{\bibinfo{volume}{60}}, \bibinfo{pages}{R29} (\bibinfo{year}{1999}).

\bibitem[{\citenamefont{Lukin et~al.}(2001)\citenamefont{Lukin, Fleischhauer,
  Cote, Duan, Jaksch, Cirac, and Zoller}}]{Lukin01}
\bibinfo{author}{\bibfnamefont{M.~D.} \bibnamefont{Lukin}},
  \bibinfo{author}{\bibfnamefont{M.}~\bibnamefont{Fleischhauer}},
  \bibinfo{author}{\bibfnamefont{R.}~\bibnamefont{Cote}},
  \bibinfo{author}{\bibfnamefont{L.~M.} \bibnamefont{Duan}},
  \bibinfo{author}{\bibfnamefont{D.}~\bibnamefont{Jaksch}},
  \bibinfo{author}{\bibfnamefont{J.~I.} \bibnamefont{Cirac}}, \bibnamefont{and}
  \bibinfo{author}{\bibfnamefont{P.}~\bibnamefont{Zoller}},
  \bibinfo{journal}{Phys. Rev. Lett.} \textbf{\bibinfo{volume}{87}},
  \bibinfo{pages}{037901} (\bibinfo{year}{2001}).

\bibitem[{\citenamefont{Saffman and Walker}(2002)}]{Saffman02}
\bibinfo{author}{\bibfnamefont{M.}~\bibnamefont{Saffman}} \bibnamefont{and}
  \bibinfo{author}{\bibfnamefont{T.~G.} \bibnamefont{Walker}},
  \bibinfo{journal}{quant-ph/0203080}  (\bibinfo{year}{2002}).

\bibitem[{\citenamefont{Greene et~al.}(2000)\citenamefont{Greene, Dickinson,
  and Sadeghpour}}]{Greene00}
\bibinfo{author}{\bibfnamefont{C.~H.} \bibnamefont{Greene}},
  \bibinfo{author}{\bibfnamefont{A.~S.} \bibnamefont{Dickinson}},
  \bibnamefont{and} \bibinfo{author}{\bibfnamefont{H.~R.}
  \bibnamefont{Sadeghpour}}, \bibinfo{journal}{Phys. Rev. Lett.}
  \textbf{\bibinfo{volume}{85}}, \bibinfo{pages}{2458} (\bibinfo{year}{2000}).

\bibitem[{\citenamefont{Boisseau et~al.}(2002)\citenamefont{Boisseau, Simbotin,
  and Cote}}]{Boisseau02}
\bibinfo{author}{\bibfnamefont{C.}~\bibnamefont{Boisseau}},
  \bibinfo{author}{\bibfnamefont{I.}~\bibnamefont{Simbotin}}, \bibnamefont{and}
  \bibinfo{author}{\bibfnamefont{R.}~\bibnamefont{Cote}},
  \bibinfo{journal}{Phys. Rev. Lett.} \textbf{\bibinfo{volume}{88}},
  \bibinfo{pages}{133004} (\bibinfo{year}{2002}).

\bibitem[{\citenamefont{Killian et~al.}(2001)\citenamefont{Killian, Lim, Kulin,
  Dumke, Bergeson, and Rolston}}]{Killian01}
\bibinfo{author}{\bibfnamefont{T.~C.} \bibnamefont{Killian}},
  \bibinfo{author}{\bibfnamefont{M.~J.} \bibnamefont{Lim}},
  \bibinfo{author}{\bibfnamefont{S.}~\bibnamefont{Kulin}},
  \bibinfo{author}{\bibfnamefont{R.}~\bibnamefont{Dumke}},
  \bibinfo{author}{\bibfnamefont{S.~D.} \bibnamefont{Bergeson}},
  \bibnamefont{and} \bibinfo{author}{\bibfnamefont{S.~L.}
  \bibnamefont{Rolston}}, \bibinfo{journal}{Phys. Rev. Lett.}
  \textbf{\bibinfo{volume}{86}}, \bibinfo{pages}{3759} (\bibinfo{year}{2001}).

\end{thebibliography}
\newcommand{\noopsort}[1]{} \newcommand{\printfirst}[2]{#1}
  \newcommand{\singleletter}[1]{#1} \newcommand{\switchargs}[2]{#2#1}

\end{document}